# Dynamic Energy Beacon: An Adaptive and Cost-effective Energy Harvesting and Power Management System for A Better Life

Nan XU, *Student Member, IEEE,* Xiao QIU, Bo XU, Junyuan SHU, Ka Ho WAN

*Abstract*—In this proposal, a cost-effective energy harvesting and management system have been proposed. The regular power keeps around 200 Watt while the peak power can reach 300 Watt. The cost of this system satisfies the requirements and budget for residents in the rural area and live off-grid. It could be a potential solution to the global energy crisis, particularly the billions of people living in severe energy poverty. Also, it is an important renewable alternative to conventional fossil fuel electricity generation not only the cost of manufacturing is low and high efficiency, but also it is safe and eco-friendly.

*Index Terms*—Energy Harvesting, Solar Panel, Power Management, Integrated Circuit Design, Power Inverter.

## I. Introduction of the "Dynamic Energy Beacon"

AFTER the 2nd Industrial Revolution, human being enters into The Electrical Era. The Electrical Generator, Stations, and Power Grid gradually emerge afterward. About one hundred years later, the invention of Integrated Circuit and World Wide Web bring us to a new stage of Information Age. The Intelligent hardware is connected via Cloud nowadays and make a Big Data World. People are enjoying the convenience of the above modern civilization.

However, there are still some areas on earth that are falling behind. According to the related reports, there are more than 3 billion people around the world are suffering from a shortage of energy, which 1/3 of them are living without power grids [1]. Their living quality, including health, education productivity and the connection with modern communities, are drastically affected. The effective solutions, which combine innovative technology and reliable products, should be addressed to these urgent and severe issues [2].

This is Dynamic Energy Beacon, a group of students from HKUST and HFUT, proposed a household level energy harvesting, regulation, operation, and storage system, with the economic viability and environmental sustainability. It consists of the following components: The Front-end Solar Panels for Energy harvesting. The central Power Management Integrated Circuits attached to the cells for electricity regulation. The Lead storage batteries are equipped as the Energy storage and the backup supply during the night. The corresponding loading applications are DC: lighting equipment (LED), as well as other AC household or consumer electronics via Inverter(s). The system diagram is shown in Fig. 1.

### A. Front-end Solar Panel

There are various types of photovoltaic materials to choose from as the solar cells. Most of them have high efficiency, while silicon solar cells are the most appropriate one with the longtime commercial application [2] [3].

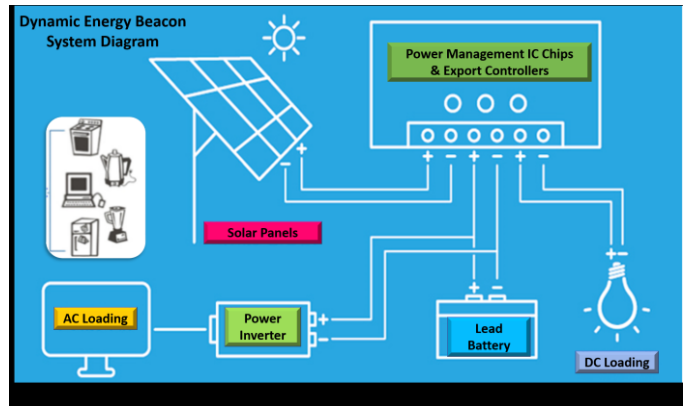

Fig. 1. System Overview of Dynamic Energy Beacon.

Recently, high-performance multi-crystalline silicon (HP mc-Si) solar cells are reported with 21.9% efficiency, which has achieved what we need [4][5]. Under the standard spectrum (AM1.5G), 1 kW/$m^2$ is given to solar panels and approximate 1.3 $m^2$ MC-Si made solar cells will provide 1000 Wh/day or 200 W peak power in high energy use family.

Another exhilarating news is that cost in the last few years have come down by a very good margin which becomes a big reason for the reduced cost of multi-crystalline solar panels. Multi-crystalline solar panels cost per 100 watts was about 40 USD now, which means that 72 panels with 300w outputting just cost as little as 120 USD [6]. In this condition, it is possible to purchase an integrated solar module under 1500 USD, which is the budget of the target audience[1]. Not only the low price but also the outstanding efficiency of the cloudy days compared with the single-crystalline solar panels, making it a large share in the market.

Optical losses chiefly affect the power of a solar cell [7]. Anti-reflection coating, surface texturing and surface passivation layers are some regular technologies to reduce optical losses and promote efficiency [8].

A regular nano-cone PDMS array can be applied on the front, which provides the functions of water-repellent and self-cleaning to the device. The structure of the nano-cone PDMS antireflection array has a depth and opening width of 1 $\mu$m (aspect ratio of 1), which effectively affects the light-harvesting capacity of solar cell devices [9]. Moreover, a drop of water is on the film with a static contact angle of 152˚ and roll-off angle of 15˚. If the dust or particles adhere to the panels, it will be flushed by rain easily just like the lotus leaves. This super-hydrophobicity adds the self-cleaning ability to the photovoltaic devices. The surface also has good mechanical properties, remaining more than 90% of the initial value even after 200 mechanical bending cycles [10]. The corresponding figures are shown in the following Fig. 2.

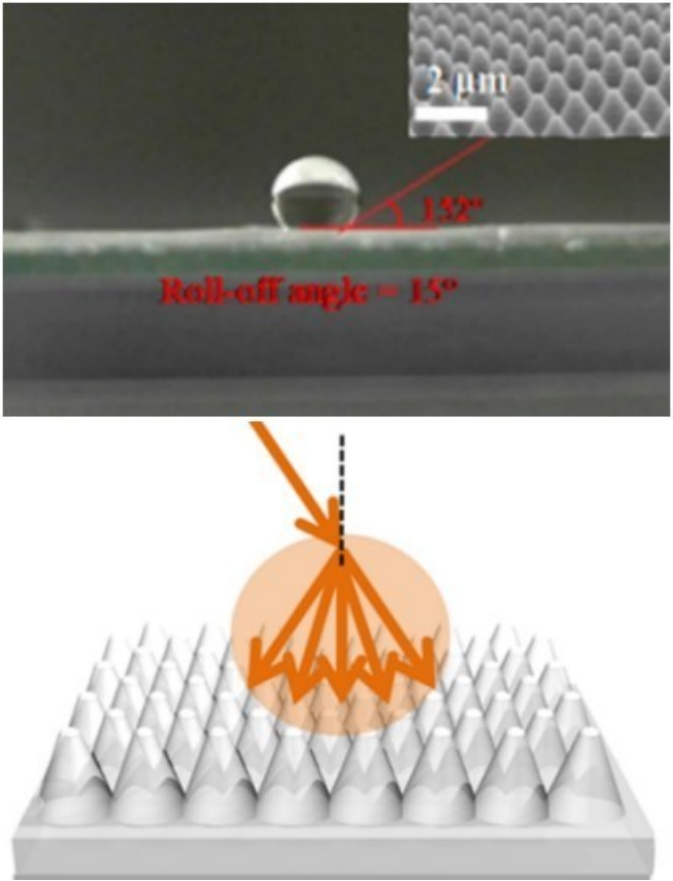

Fig. 2. (1)Photograph shows a drop of water on the surface of the nanocone PDMS array with static contact angles of 152° and roll-off angle of 15°. (2)The cones with a depth and opening width of 1 $\mu$m (aspect ratio of 1), which can improve the light-harvesting capacity of solar cell devices.

As for the production process, the low pressure and temperature thermal oxidation process could be an efficient way to increase the efficiency of the multi-crystalline silicon solar cells[10].

By deploying the above-advanced processing technologies, the solar module has a service life of more than ten years and it is working normally from -40°C to 85°C. More detailed parameters can be found in the Appendix.

*B. Power Management ICs for the Electricity Regulation*

According to the diverse characteristics of different sunlight cases and loading conditions, a dedicated Power Management Integrated Circuit (PMIC) has been proposed for the whole Dynamic Energy Beacon system. The prototype of this product is from[11]. The PMIC basically is a boost inductive DC-DC converter consists of the large power transistors, an off-chip inductor, an output capacitor and the control stage, the low power digital circuits to define the operation modes.

The converter is equipped with the feedback ability to clamp the Photo-voltaic cells to the optimized point where the maximum solar power is delivered. Meanwhile, the DC-DC converter, which sources the energy from the PV Cell, also provides a stabilized output power to the loading circuits or systems. Furthermore, the additional digital logics will help the scheduling of energy utilization among different front-end power sources and back-end electronics.

If the external power is sufficient and relatively constant, the PWM (Pulse Width Modulation) mode will be activated which the PMIC will not only do the power regulation but also the charging work to the backup battery. If the power from the solar panel is not enough, then the converter will enter into the PFM (Pulse Frequency Modulation) mode, and not only keep on sourcing the power from the source, but also absorb the power from the backup battery, as well as the management of loadings to prolong the lifespan of the overall system. These two modes are shown in Fig. 3.

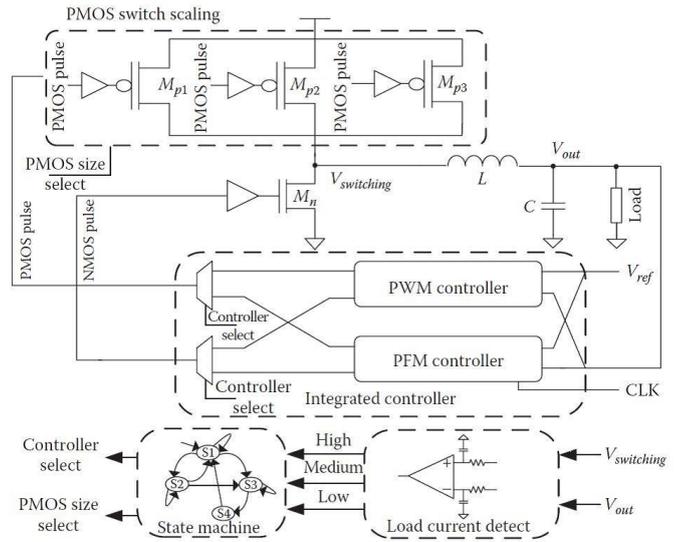

Fig. 3. System Diagram of DC-DC Converter.

When the PMIC is working, it can adjust its input impedance through tuning the switching clock frequency. The power is then delivered to the loading applications directly, or to a rechargeable backup battery for the night or emergency use. The details of the IC, simulation/measurement results, and table of specifications can be found in the Appendix.

For safety reasons, an overvoltage protection (OVP) and over-temperature protection (OTP) function are built into the I/O pins of the energy source as well as the interfaces of the PMIC and other parts. In conclusion, the PMIC acts as a good power manager connecting the Front-end Solar panel, Energy Storage and loadings for long-term use.

*C. Energy Storage for Backup use*

Valve-regulated lead-acid battery (VRLA battery) can be considered as the energy storage of solar cell modules. According to the off-the-grid system and climate conditions in specific areas, VRLA can be packaged into sealed boxes, buried underground and connected with other modules via cables. In such a solution, it could be quite area-saving and safe for users, as well as fewer requirements for maintenance. The parameters of the standard VRLA battery[12] can be found in the Appendix.

## D. Power Inverters for DC/AC conversion

The general power system consists of power plants, transmission links, power transformation links, distribution links, and power users. Due to regional reasons, power generation is sometimes far away from power users the substation and distribution links that are often accompanied by loss of energy and huge losses are very important.

There are many advantages to using Power Inverter. First, it is convenient to converts DC power into AC types of power. Second, the DC transmission amount of energy is around 1.5 times of that in AC. Last but not least, the Eddy current of AC cables can be drastically decreased when using DC transmission, which means the improvement of power quality. This comparison of DC and AC powering is shown on the right-hand side Table I.

| Advantages of DC Powering | Explanation and Comparison with AC |
|---|---|
| Increasing of the Power Capacity with the same cross-section and current density | $P_{DC}/P_{AC} = \frac{2 \times V_{DC} \times I_{DC}}{\sqrt{3} \times V_{AC} \times I_{AC} \times \cos(0.9)}$ $= \frac{2 \times (\sqrt{2} \times V_{AC}) \times (I_{AC}/\sqrt{3})}{\sqrt{3} \times V_{AC} \times I_{AC} \times \cos(0.9)} \approx 1.52$ |
| Improvement of the Power Quality | The Power Inverter can flexibly emit or absorb the Reactive power, as well as the Compensation and Regulation |
| Reduction of Energy Loss | The eddy current effect and phase shift in AC can be eliminated in DC cases |

TABLE I
THE COMPARISON OF PACKAGING TECHNOLOGIES.

## E. Packaging Technology

The packaging technology has been scaled down from macro PTH (Pin Through Hole) to SMT (Surface Mount Technology) [13] to AAP (Area Array Package) [14] [15]. The characteristics and comparison among these 3 generations are shown in the following Table II. Regarding the realistic cost and quality requirement of our Dynamic Energy Beacon system, the mature SMT can be applied to attach the Power Management IC Chips onto the backside of Solar Panels.

| Packaging Technologies | Typical Categories | Characteristic | Advantages | Drawbacks |
|---|---|---|---|---|
| First Generation | PTH (Pin Through Hole), including DIP, TO, PGA. | Connecting to PCB board through holes directly. | Simple; easiest way to connect with PCB board. | Difficult to increase the density and frequency; hard to meet requirements of efficient mass production. |
| Second Generation | SMT (Surface Mount Technology), including SOP, PLCC, QFP, LCCC. | Mounting the chips on the PCB with tiny leads. | Package density is improved as leads are thin and short; smaller size, lighter weight; easy to mass production. | The number of I/O pins and Operation frequency of the circuits are not high enough. |
| Third Generation | AAP (Area Array Package), including BGA, CSP, Flip-Chip. | The pins are replaced by solder balls. | Superior electrical and thermal performance; high I/O pin count; smallest size. | Complicated technology; high cost. |

TABLE II
THE COMPARISON OF PACKAGING TECHNOLOGIES.

## II. HIGHLIGHTS OF THE SYSTEM

### A. Impact Highlights

Our product is focusing on Track 1A, which is basically a solar home system dedicated to single-family use at the Tier 2 level without creating other redundant infrastructure in advance when it is needed.

It is quite easy to use which the solar panels are supposed to be installed on the roof or the side-walls wherever the sunlight can be sufficiently absorbed, the Power manager Chips are attached to the panel for the precise regulation and management, the VRLA batteries are safely buried underneath while a few cables are used to connecting every module into a whole system.

Compared with the conventional off-grid energy system, the main difference is replacing the Macro bulky controller with the Micro Integrated Circuits. Not only the volume is drastically decreased, but also the safety and long-term use are assured. For example, it is troublesome and difficult to fix when the traditional controller is not working, which leads to the whole system shutting down. If only one or two chips fail, the rest of them can still work properly, and all the necessary maintenance is to replace the failure chips with inserting new ones, which is quite easy for users. Besides, the advanced processing technologies of the Solar cell can drastically improve the light absorption and energy conversion ratio, as well as prolong the lifetime of the whole panel.

The whole system of Dynamic Energy Beacon is equipped with user-friendly property: The installation of Solar panels is quite easy without any sophisticated skills, just open the panels and fix them with screws. The Power management ICs are attached to the backside of the Solar panel, which the output power can be delivered through cables to either the indoor equipment/power inverters, but also outdoor lead batteries. What's more, the bulky batteries can be either put indoor or buried underground freely by users as well. Last but not least, customers can define the combination of above all the components as they wish, to satisfy the actual needs.

The total cost of one system is around 240 USD and once it is installed, it can function well for around 3-5 years at least. It is affordable and less than 1/5 of the whole budget, if we take 3 years budget into consideration, to the target audience living in particular areas. Besides, it is portable since the aforementioned 3 major components are not that bulky. It is easy for users to bring and re-install whenever they need to migrate from one place to another.

### B. Technical Highlights

Due to the multiple solar cells used, the peaking power absorbed externally can reach 600 W, and normally output around 400 W. Taking the energy loss of conversion and the efficiency of Power management Chips, still, we can assure around 200-300 W electricity power delivered to the household equipment. Since most energy is from the daytime sunlight, which may last around 6 - 10 hours on average, the total power absorbed can be around 1.8KWh to 3KWh. It is enough for both day and night use for a single-family.

The expandability of the product is obvious, which means we can sourcing more energy by simply adding and plugin more solar panels and batteries. The use of the whole system is quite easy without much more advanced knowledge or training. The localization and distributed properties make it easily spread.

## C. Business Highlights

There are some similar products existing on the market: Hanergy Thin Film Power [16] and Indoor mini generator. The detailed comparison of them and Dynamic Energy Beacon are as follow Table III:

| | Electricity Power Output | Cost (USD) | Remark (Pros & Cons) |
|---|---|---|---|
| Hanergy Thin Film Power | 90-105 W | 20/$m^2$ | Mechanical Load: 2.4 KPa<br>Expected Lifespan: 3-5 Years |
| Indoor mini generator | 3000 W | 73 (Generator) + 20 / (15kg Gas) | Hard for operation/maintenance<br>Only for backup use<br>Not environmental friendly |
| Dynamic Energy Beacon | 200 W (Usual)<br>300 W (Peak) | 60/$m^2$ | Mechanical Load: **5.4 KPa**<br>Expected Lifespan: **5-8 Years**<br>**User Friendly** and **Expandable** |

TABLE III
COMPARISON WITH EXISTING PRODUCTS ON THE MARKET.

The potential customer can be divided into 2 channels, with respect to the basic and advanced version of our products.

*1) Basic Version:* Targeting to single-family in energy poverty areas. Each block is $2m^2$ which consists of 2 pieces of Solar panels, attached power management ICs, power inverter, and 2 large volume lead batteries. It can provide 300W at peak time, which is mainly for simple daily household electric appliances, as well as the evening lighting requirements.

*2) Advanced Version:* Targeting public facilities and government offices. Larger block sizes with more components included. At most 2.5kW power can be provided per $20m^2$. Apart from components in 1), the high-performance illuminating coating materials [23] [24], as indicated in Fig. 4 can be applied to some public areas such as stair landings, exit passages or underground tunnels. Only a few LEDs can be the light source, and duration of illumination can last more than 10 hours.

The price for the whole system is 240 USD per set for the Basic version. The detailed costs are shown in Table IV.

| Item | Price (USD) | Quantity | Amount (USD) | Lifetime |
|---|---|---|---|---|
| Solar Panel | 90 | 2 | 180 | 5-10 years |
| Power Management IC | 0.8 | 25 | 20 | 5-10 years |
| Lead-acid (VRLA) Battery | 10 | 2 | 20 | 3-5 years |
| Peripheral equipment | NA | NA | 10 | 3-5 years |
| Shipping and Installation | NA | NA | 10 | NA |
| Maintenance | NA | NA | Free within 3 years<br>50 from next year | NA |
| Total | NA | NA | 240 (Additional 20 from the 4th year) | NA |

TABLE IV
THE COSTS OF ALL THE COMPONENTS AND LIFETIME.

We will focus on the selling of Advanced versions at the beginning, which is customized according to the requirements from the government, public facilities and major companies at the target area. The price will be negotiable, with special discount or add-value service provided, to reach the win-win to both parties. After the market opening from high-end customers, the sells of Basic version will spread among individual local residents or families. Due to the low population density in particular areas, say, for example, 20-1000 per community, a special offer with 10% discount or free installation tutorial can be applied to accumulate the original users.

With the selling at 2 parallel channels, the revenue is from 2.4K USD to 120K USD for the first year. While with the promotion from ourselves, the collaborators, as well as the users themselves, our revenue would be drastically increased soon. The major expenditure is fabrication, logistics, and local sales. The Initial Capital would be around 100K USD for the first year. When the project goes to the second year, it could reach our break-even date and start profiting from the 3rd year. The details can be found in Fig. 10 in the Appendix.

The potential collaborator and investors, which are summarized in Table VII in the Appendix, can be Huawei [17], China Electronic Systems Technology Corporation [18], GLOBAL FOUNDRIES [19], Renewit [20], All Solar [21], Power- Solution [22] and DHL, TNT Express, also the local governments and communities for their subsidy and promotion.

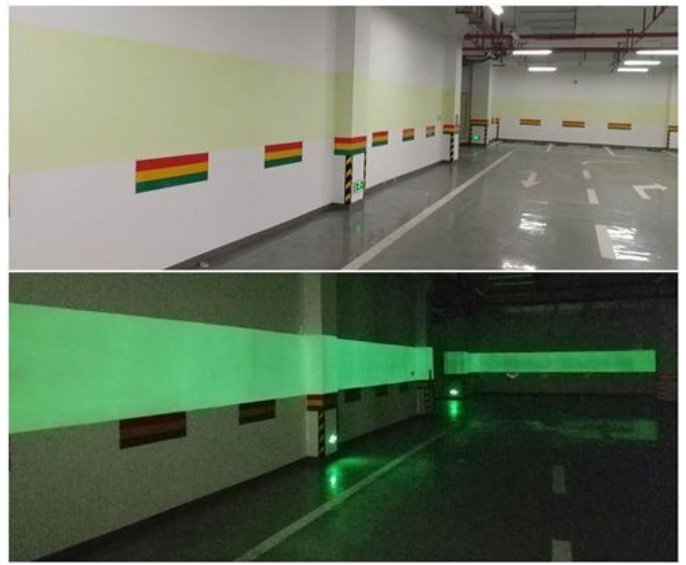

Fig. 4. Comparison of illuminating material in light and dark conditions

Besides, above all are only the features of the First generation of our products. With more advanced and iterated technologies, later on, the upgrading about improving the conversion efficiency by increasing the Transmittance, hybrid combination of Multi-crystalline Si and Single-crystalline Si adaptive to the weather condition, enhancement of peak power as well as the multi-function by optimization of Power Management ICs, will be conducted on a regular of 2 years.

III. PROPOSED FIELD TESTING PLAN

*1) Identification of Target Community:* The target community which we will choose is the remote village in Yunnan Province, China. There are many optional places [25][26][27] in China that are corresponding to the power purchasing requirement. The detailed comparison of them will be given in the following Table V, and the reasons for our decision are:

| Area | Latitude | Year Average Temperature (°C) | Per Capita Disposable Income (CNY) | Amount of Precipitation (mm) | Annual Solar Radiation Amount (kWh/$m^2$) |
|---|---|---|---|---|---|
| Yunnan, Xishuangbanna | 21°08'-22°36' | 20 | 12043 | 1193.7-2491.5 | 1393-1625 |
| Gansu, Jiuquan | 38°09'-42°48' | 3.9-9.3 | 15764 | 84 | 1625-1855 |
| Heilongjiang, Daxinganling | 50°10'-53°33' | -2.8 | 12096 | 746 | 1163-1393 |

TABLE V
COMPARISON OF THREE CITIES IN MAINLAND CHINA

The latitude and average temperature of Xishuangbanna are most similar to those of Africa. PCDI is the lowest one of the three cities. With high temperatures and a large amount of rain, the tropical rainforest climate here is a typical match. Taking these into consideration, the late Spring and early Summer will be a good time to possess field tests in Xishuangbanna.

*2) Field Testing Plan and Schedule:* Combining our product and the local conditions, we propose a detailed plan which could be efficiently carried out in the following Table VI.

| Steps | Time Cost | Details |
|---|---|---|
| First | Before departure | Contacting with local government and communities to apply an introduction letter to provide our reliability towards target homes. |
| Second | 2 days | Visit chosen villages, learning their power usage habits and inquiring their intentions of trying out our product one home by one. |
| Third | 1 day | Selecting 10-20 eligible homes from last step and Drafting on-site plans for each home to install our product. |
| Fourth | 2 days | Installing the products and independent data acquisition (DAQ) systems in selected homes. Teaching users the usage instructions. |
| Fifth | 7 days | Collecting data from DAQ and analyzing them. Preparing for any unforeseen circumstances (system out of work, broken solar panel, damage chips, etc.) and recording for product improvement. |
| Sixth | 2 days | Recycling our products and enquiring these homes for both user feedback and their willingness of purchasing our system. |

TABLE VI
PLAN AND SCHEDULE FOR FIELD TESTS

*3) Field Test Goals and Evaluation Criterion:* In this field test, our goals including two aspects. One is learning the performance of our product in practical situations so that we can improve it targeted. Another one is finding out the attitude of people living in a remote area towards our product, then we could learn their actual needs and the prospect of product. During the survey time, we can learn about the needs of local people on power usage. And the running data of our system can be collected in the testing period. Lastly, in the feedback round, we will know our product is welcome or not among target homes. To evaluate our goals, we could judge the performance from data collected by the DAQ system and the error reports from target homes. If the whole system works normally most of the time and the error reports are less, we could consider our system having the ability to work in practical situations. And if most of the target homes are willing to purchase our system, it proves the product having commercial potential.

*4) Statement of Risks:* To avoid the possible extreme weather such as typhoons, we choose Yunnan Xishuangbanna which is an inland city. In order to get more similar data as Africa, we will arrange our field test to be carried out in early summer so that the sunlight condition and temperature are proper. As for the avoidance of the risk of product, we will fix the solar panel tightly and high enough and the equipment of electric cables will be carefully designed.

IV. CONCLUSION

By integrating high-efficiency Solar panels, Power management ICs, Lead batteries, and Inverters with mature packaging technology, the "Dynamic Energy Beacon", a cost-effective Energy harvesting and power management system, has been proposed. After several primary measurements of prototypes, the estimated output electricity power keeps around 200W (Peak at 300W). It could be a potential solution to the billions of people living in severe energy poverty. The cost of this system satisfies the requirements and budget. Besides, it is an innovative renewable alternative to conventional fossil fuel not only the prices of all the modules, packaging and installation are low, but also it is safe and Eco-friendly.

V. APPENDIX

The group members are postgraduate and undergraduate students from The Hong Kong University of Science and Technology (HKUST) and Hefei University of Technology (HFUT) respectively.

The workload is distributed as follows: Nan XU is the team leader, and responsible for the whole idea, as well as the team member recruitment, testing of Power Management ICs and the final proposal integration. Xiao QIU is responsible for the investigation of the front-end solar cell/panel. Bo XU is responsible for the design of the proposed DC-DC converter, simulation, debug, Packaging investigation and Field testing plan. Junyuan SHU is responsible for the review of power inverter part for the DC/AC conversion and Measurement of Solar cell samples. Ka Ho WAN is responsible for the market and business plan drafting.

The corresponding figures and tables are on the next page.

ACKNOWLEDGMENT

The authors would like to express the sincere gratitude to Prof. KI Wing-Hung from Integrated Circuit Design Center (ICDC), Prof. FAN Zhiyong from Functional Advanced Nanostructures Laboratory (FAN Lab) of HKUST, and Beijing Lingwu Science and Technology Co., Lt for the authorization of Intellectual Properties used in our proposal. The authors would also express special thanks to Ms. Jenny Chan from HKUST for her precious advice and suggestions.

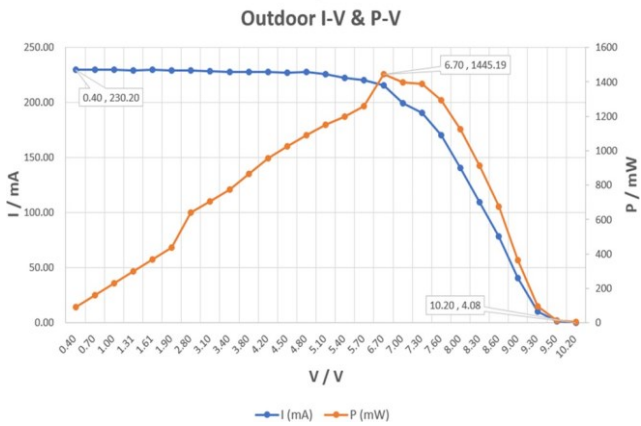

Fig. 5. Outdoor Test Results of Solar Cell Sample

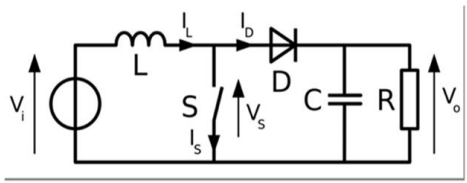

Fig. 6. Structure of the Boost Converter

| Items | Simulation results |
|---|---|
| Technology | 0.35 CMOS Process |
| Vg / Vo / Ripple | 1.2V / 2.4V / 4.45mV |
| Frequency | 30 MHz |
| Loading Current | 100mA to 500mA |
| Efficiency | 70% - 85 % |

TABLE VII
THE SIMULATION RESULTS OF DC-DC CONVERTER

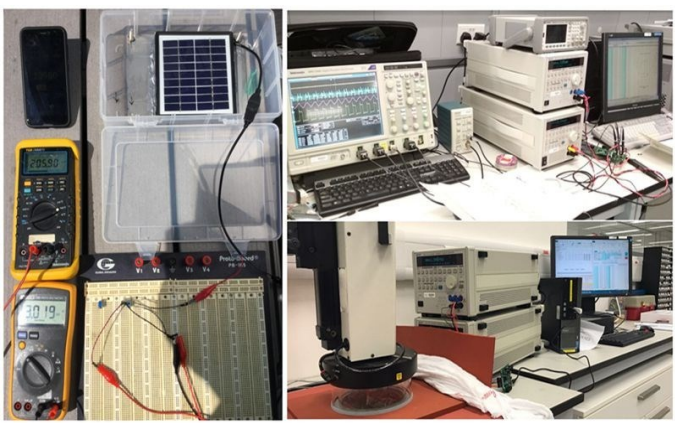

Fig. 7. Function and Thermal Testing setup of Solar Cell Sample and DC-DC Converter

| Equipment | Function |
|---|---|
| ADCMT 6243 | Loading Current |
| ADCMT 6244 | Voltage Source |
| DPO 7104C | 4 Channel Oscilloscope to monitor the output |
| Agilent E3610A | Power supply to Power Management IC (Will be replaced by Solar Panel) |
| Fluke 15B | Portable Digital Multi-meter for debug |
| Thermo-Stream | Heat source for the Temperature test of Power Management IC |

Fig. 8. List of the testing equipment for Power Management IC

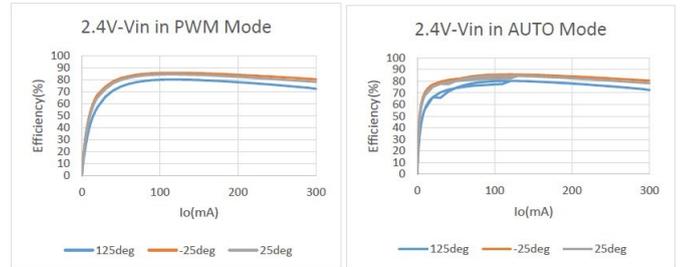

Fig. 9. Efficiency measurement in different Temperature

| Item | Solar Panel | VRLA Battery |
|---|---|---|
| P_max | 300W | 200Ah(Capacity) |
| V_Max/ V_oc | 36V / 43V | 12 V |
| I_mp/ I_sc | 8.33A / 9.17A | 2.4KWh(Energy) |
| Temperature | -40℃ - 85 ℃ | |
| Size/Weight | 2m*2m*50mm/24kg | 60kg |

TABLE VIII
THE PARAMETERS OF SOLAR PANEL AND VRLA BATTERY

| Company / Institute | Role of Collaboration |
|---|---|
| Huawei | Base Station and Advertising |
| China Electronic System Technology Cooperation | Packaging and Supply Chain |
| GLOBAL FOUNDRIES | Power IC fabrication |
| Renewit | Solar Panel |
| All Solar | Solar Lighting |
| Power-Solution | and Inverter |
| Local Government / Communities | Advertising Promotion |
| DHL / TNT Express | Logistics |

TABLE IX
THE POTENTIAL COLLABORATORS AND THEIR ROLES

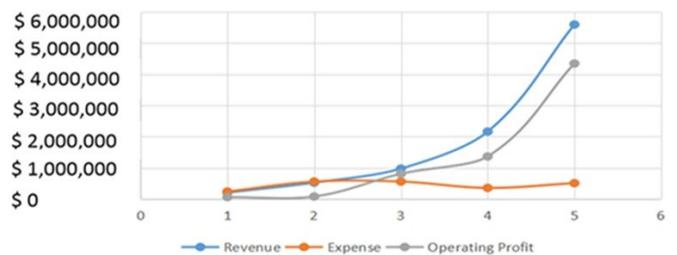

Fig. 10. Yearly Profit and Loss Statement

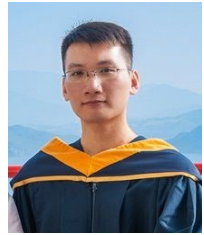

**Nathan, Nan XU** (S'17) received his B.ENG. degree in Electronic Science and Technology from Hefei University of Technology (HFUT), Hefei, Anhui, China, the M.Sc. degree in IC Design Engineering and M.Phil. in Electronic and Computer Engineering from the Hong Kong University of Science and Technology (HKUST), Hong Kong.

He is currently pursuing the P.h.D Degree in Individualized Interdisciplinary Programs (IIP) at HKUST. After joining the Integrated Power Electronics Laboratory (IPEL), which now extended to Integrated Circuit Design Center (ICDC), and before the M.Phil. study, he was a Research Assistant working for the CERN-ATLAS-HK project and collaborating with the staff from HKU, CUHK, and USTC. His current research interests include Power Management Integrated Circuit Design and Test.

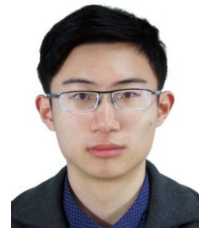

**Xiao QIU** is currently pursuing a Bachelor's Degree in Material Physics at Hefei University of Technology (HFUT) Hefei, Anhui, China. He has worked on nanocluster at Hefei Institute of Physical Science, Chinese Academy of Sciences. After joining the Anhui Key Laboratory of Advanced Functional Materials and Devices and working for Lithium-Ion Battery, he continued his work in Nanoscience and Nanodevices Lab (NNL) at National Tsing Hua University (NTHU). His current research interests include Energy Devices and Materials.

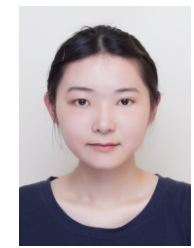

**Bo XU** received her Bachelor degree in Opto-electronic Information Science and Technology from Huazhong University of Science and Technology. She received her M.Sc. degree in IC Design Engineering from The Hong Kong University of Science and Technology, where she is pursuing her PhD degree in ECE with focus on visible light communication..

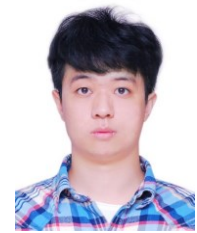

**Rocky, Junyuan SHU** received his B.ENG. degree in Automation from South China University of Technology (SCUT), Guangzhou, Guangdong, China. He is currently pursuing an M.Sc. degree in IC Design Engineering from the Hong Kong University of Science and Technology (HKUST), Hong Kong. Before the M.Sc study, he participated in the Free-scale contest and won the Excellent Team Award. He also got prizes in Intelligent Control Competition in SCUT. He has been interned in China Administration of Power Supply, He Gang Branch.

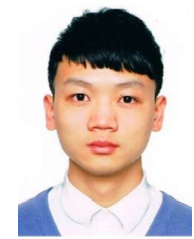

**Terrance, Ka Ho WAN** is now pursuing his undergraduate degree in Operation Management and Economics at Hong Kong University of Science and Technology. He is a member of Alpha Kappa Psi, the oldest and biggest business fraternity in the world. He received the honor of Hong Kong Top 10 Outstanding Youth (Open Category, aged 19-28) jointly organized by COY and HAB at the age of 19. He joined the entrepreneur events organized by the World Trade Organisation, Tsinghua University, and the likes.